# C-Band VSAT Data Communication System and RF Impairments

T.P. surekha [1]   T. Ananthapadmanabha [2],   C. Puttamadappa [3]

[1] Research Scholar at NIE and Associate Professor, Dept. of E&CE, Vidyavardhaka College of   Engineering, Mysore, India. *E-mail: tpsuriramesh@gmail.com*

[2] Professor, Dept. of E&EE, National Institute of Engineering,  Mysore, India And Honorary secretary of  IEI, Mysore local center, Mysore, India.
*E-mail : drapn2008@yahoo.co.in*

[3] Professor, Dept of E&CE, Sapthagiri college of Engineering. Bangalore, India.
*E-mail: puttamadappa@gmail.com*

## ABSTRACT

*This paper is concerned with modelling and simulation  of VSAT (very small aperture terminal) data messaging network operating in India at Karnataka with extended C-band. VSATs in Karnataka of KPTCL use VSATS 6.875-6.9465G Hz uplinks and 4.650- 4.7215 GHz downlinks. These frequencies are dedicated to fixed services. The Satellite is Intelsat -3A, the hub has a 7.2 m diameter antenna and uses 350W or 600W TWTA (Travelling wave Tube Amplifier). The VSAT's are 1.2 m with RF power of 1W or 2W depending on their position in the uplink beam with data rate of 64 or 128 K bit/s. The performance of the system is analysed by the error probability called  BER (Bit Error Rate) and  results are derived from Earth station to hub and hub to Earth  station using satellite Transponder as the media of communication channel. The Link budgets are developed for a single one-way satellite link.*

## I. INTRODUCTION

This paper is concerned with VSAT (Very Small Aperture Terminal). VSAT is the main communication media for Karnataka Power Transmission Corporation Limited (KPTCL)/ ESCOM (Electric Supply Companies) Supervisory Control And Data Acquisition (SCADA) network. It also provides voice communication to all KPTCL, ESCOMS stations and major generating stations with load dispatch centre (LDC). Implementation deals with the modelling and simulation of  RF  communications link involving satellite transponder. A transponder is a series of interconnected units forming an RF (Radio Frequency) broadband communication channel between the receiver and transmit antenna in a communication satellite. The typical extended C-band communications satellite will be examined briefly here. Each transponder is amplified by either a travelling wave tube amplifier (TWTA) or a solid state power amplifier (SSPA). Satellites of this type are very popular for transmitting TV channels to broadcast stations, cable TV systems (DTH) direct to home systems. Other applications include Very Small Aperture Terminal (VSAT) data communications network. Integration of these information type is becoming popular as Satellite transponders can deliver data rates in the range of 1000 Kbps  to 256 Mbps. Achieving these high data rates require careful consideration of the design  and performance of the repeater. The methodology adopted here is QAM technique and software-tool is Math works with simulink to explore the end to end simulation of communication links involving satellite transponder.

Fig1shows the block diagram of basic VSAT satellite communication system. Satellite communication system consists of many earth stations on the ground and these are linked with a satellite in space. The user is connected to the Earth station through a terrestrial network and this terrestrial network may be a telephone switch or dedicated link to earth station. The user generates a baseband signal that is processed through a terrestrial network and transmitted to a satellite at the earth station. The satellite transponder consists of a large number of repeaters in space which receives the modulated RF carrier in its uplink frequency spectrum from all the earth station in the network, amplifies these carriers and retransmits them back to the earth stations in the downlink frequency spectrum. To avoid the interference, downlink spectrum should be different from uplink frequency spectrum. The signal at the receiving earth station is processed to get back the baseband signal, It is sent to the end user through a terrestrial network.

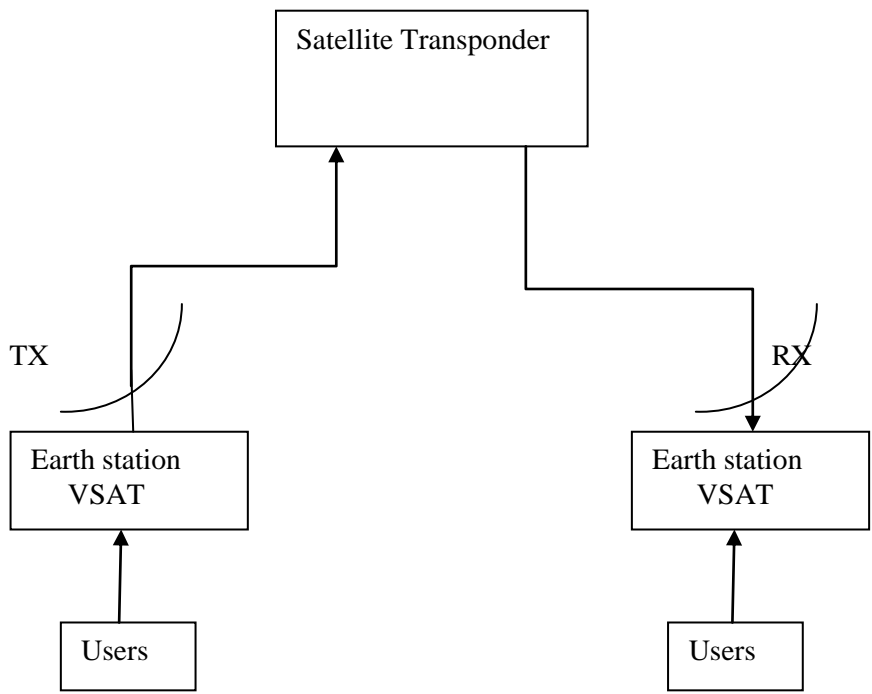

FIG 1: Basic VSAT satellite communication system

Commercial communication satellite uses a frequency band of 500 M Hz bandwidth near 6 G Hz for uplink transmission and another 500 M Hz bandwidth near 4 G Hz for downlink transmission .An uplink of 5.725 to 7.075 G Hz and downlink of 3.4 to 4.8 G Hz is used. Here the extended C-band is used with uplink frequency of 6.9350-6.9465 GHz and downlink frequency of 4.710-4.7215 GHz downlink frequency. Modulation used here is QAM to save the bandwidth. Extended C band is most popular because of less propagation problem. Rain attenuation and sky noise is low at 4 GHz downlink frequency of C band. So it is possible to build a receiving system.

The basic block diagram of an VSAT earth-station Transmitter is as shown in fig 2. The baseband signal from the terrestrial network is processed through modulator and then it is converted to uplink frequency. Finally it is amplified by high power amplifier and directed towards the appropriate part of antenna. The block diagram of an VSAT earth station receiver is as shown in fig 3.

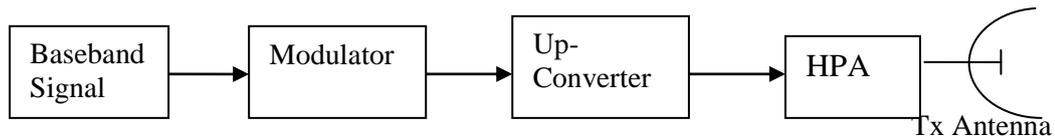

FIG 2: Block diagram of VSAT Earth station Transmitter

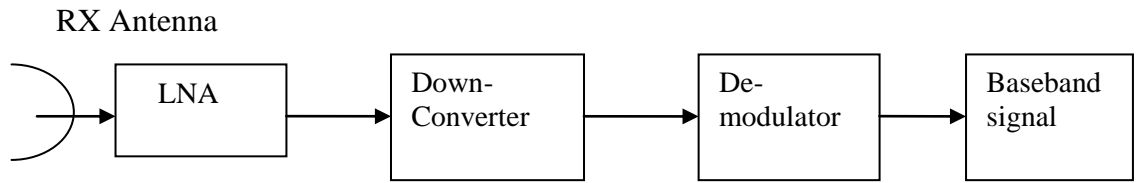

FIG 3: Block diagram of VSAT Earth station receiver

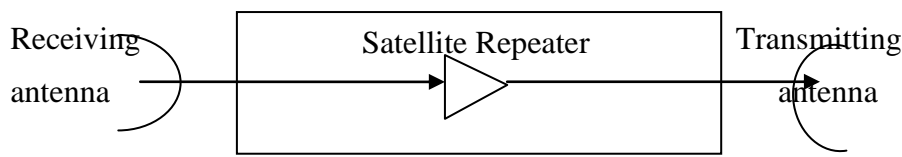

FIG 4: Block diagram of Satellite Transponder

The signal received from the satellite is processed through LNA (Low Noise Amplifier). Then it is down-converted and demodulated. Thus the original baseband signal is obtained.

## II. VSAT

The acronym VSAT is the earth station antenna used at the earth stations. In VSAT the earth station antenna size is typically less than 2.4 m in diameter and the trend is towards even smaller dishes measuring not more than 1.8 m in diameter. According to European Telecommunication standard Institute, VSAT is referred as satellite transmit –receive system that has an aperture size smaller than $2.8m^2$. VSAT's provide cost effective solutions for the growing telecommunication needs through- out the world. Today's satellites are more powerful, enabling the use of smaller and less-expensive antennas on the ground. Also, the developments include most of the necessary VSAT functions, which makes VSATs more effective.

The architecture of the networks is of two types. One is star topology and the other is mesh topology. The star topology is the traditional VSAT network topology. Here the communication link are between the hub and the remote terminal. This topology is well suited for data broadcasting or data collection. This is not applicable for speech services because the time delay is too severe (500ms). The access techniques used in a star network can be both FDMA and TDMA. In mesh topology there is a direct communication between the remote

VSAT terminals. This minimizes the time delay which is concerned with speech services. The access method used in mesh network is FDMA. Very Small Aperture Terminals (VSATs) are designed for data transmission and distribution over a wide geographical area amongst a large number of locations. The small size and low transmit power of a VSAT station are the factors that keep the price of the earth station at a level that makes a VSAT network an economic alternative to a terrestrial data network using telephone lines and modems. The hub usually houses a central host computer, which can act as a data switching centre. The architecture of the network naturally becomes star shaped, [*Maral, 1995*]. The links from the hub to the VSAT are called *outbound* links. The links from the VSAT to the hub are called *inbound* links. Both inbound and outbound links consist of two parts, uplink and downlink. It is not unusual that inbound and outbound links operate at different transmission speeds, i.e. in *asymmetrical* mode.

**2.1. Simulation model of complete system**

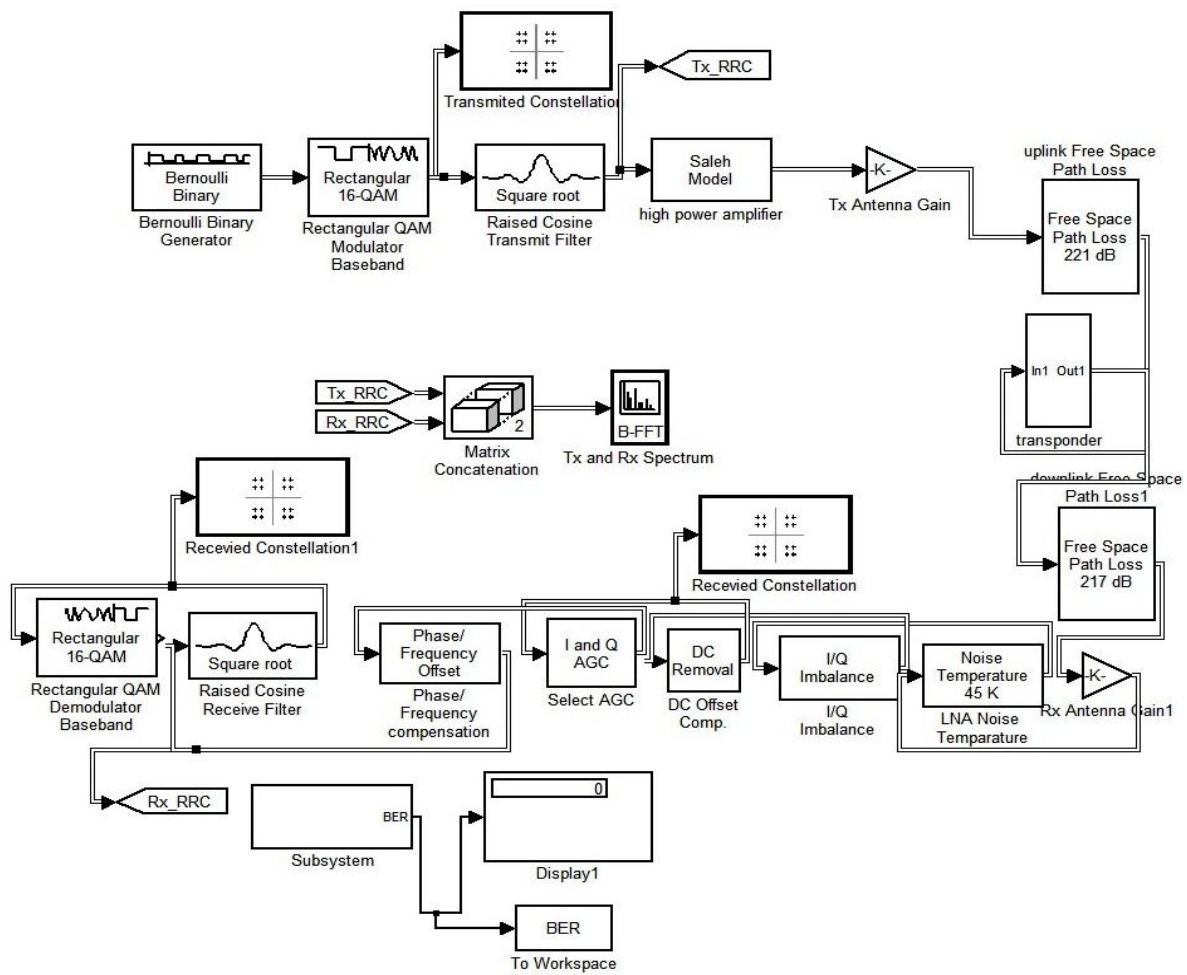

FIG 5: VSAT Satellite transponder link model

## 2.1 Model of Transmit Earth Station subsystem

1. **Bernoulli Binary generator:** This block generates random binary numbers using Bernoulli binary distribution. Data signal generates a stream of information bits to be transmitted by the transmitter, Specifically Bernoulli binary generator is employed as a data signal generator.

   **Functional Block Parameters -**
   Sampling time: 4/100000s
   Frame-based outputs: Samples per frame: 512

2. **Rectangular QAM Modulator Baseband:** The Rectangular QAM Modulator Baseband block modulates using M-ary quadrature amplitude modulation with a constellation on a rectangular lattice. The output is a baseband representation of the modulated signal. The Rectangular QAM Modulator Baseband block provides the capability to visualize a signal constellation from the block mask.
   **Functional Block Parameters -**
   - M-ary number : 16
   - Constellation ordering : Gray
   - Minimum distance (Distance between two nearest constellation points): 2

3. **Raised Cosine Transmit Filter**: The Raised Cosine Transmit Filter block up samples and filters the input signal using a normal raised cosine FIR filter or a square root raised cosine FIR filter.
   **Functional block Parameters -**
   - Filter type : Square root
   - Roll off factor : 0.2
   - Up sampling factor: 8

4. **High Power Amplifier**: Satellite links have limited power available to them. Efficient amplifiers are necessary to use this power carefully. So high power amplifiers are used. Travelling Wave Tube Amplifier (TWTA) is used to amplify radio frequency signals to high power. Model of a travelling wave tube amplifier (TWTA) using the Saleh model is used. The saleh model provides an independent gain and phase function that has been used to compute output wave-form for given input which is not available in other high power amplifications.
   **Functional block parameters -**
   - input scaling factor: −16.1821
   - AM/AM parameters [alpha beta] : [2.1587  1.1517]
   - AM/PM  parameters[alpha beta] : [4.0033  9.1040]
   - Output scaling factor:  32.9118

5. **Gain (Tx. Dish Antenna Gain):**  This block models the gain of the transmitting parabolic dish antenna on the satellite. A parabolic antenna is an antenna that uses a parabolic reflector, a surface with the cross-sectional shape of a parabola, to direct the radio waves. The main advantage of the parabolic antenna is that it is highly directive.

   Gain = 52.48 dB

## 2.2 RF satellite Transponder

Simulink model of RF satellite transponder subsystem is shown in Fig 6. Following are the components of the RF satellite link subsystem:

1. **Uplink path:** Uplink connects transmit earth station to satellite. Free Space Path Loss is the major loss that can occur during transmission to satellite. So Free Space Path Loss block is used to model this loss. This block aattenuates the signal by the free space path loss.
   Uplink free space loss: 221 dB

2. **Transponder**: A transponder receives, amplifies and transmits radio signals at different frequency. After receiving the signal a transponder will broadcast the signal at a different frequency.
   Sat. Tx. Ant. Gain: 31dB
   Sat. Ant. Rx Gain: 38.2dB

3. **Downlink path:** Downlink connects satellite to the receive earth station. In addition free space path loss Doppler and phase errors are also modeled in this path. Free Space Path Loss block attenuates the signal. Phase / Frequency Offset block rotates the signal to model phase and Doppler error on the link.
   Downlink path loss: 217dB

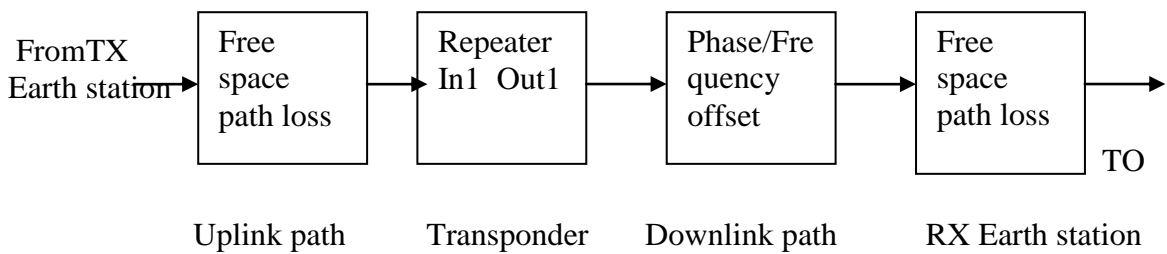

FIG 6: Model of RF satellite Transponder subsystem

## 2.3 Receiver Earth Station subsystem

Lower half of fig 5 shows the model of Receiver Earth Station subsystem. Following are the components of the receiver earth station subsystem:

1. **Gain (Rx. Dish Antenna Gain) :** This block models the gain of the receiver dish antenna( feed prime focus antenna used in VSATs)
   Gain = 36.85 dB

2. **Receiver Thermal Noise:** This block adds noise that represents the effective system temperature of the receiver.
   **Functional block parameters:**
   Specification method: Noise temperature
   Noise temperature (K): 45 K

3. **I/Q imbalance:** In I/Q (In phase/Quadrature phase) processing receivers, matching the amplitudes and phases of the I and Q branches is of major concern. In practice, matching can never be perfect in the analog front-end which results in insufficient rejection of the image frequency band. I/Q imbalance block is used to model this effect. This block creates a complex baseband model of signal impairments caused by imbalances between in-phase and quadrature receiver and aplies an in-phase dc offset, specified by the I dc offset parameter, and a quadrature offset, specified by the Q dc offset parameter, to the signal.
   **DC offset compensation:** This block is built to estimate and remove the DC offset from the signal. It compensates for the DC offset in the I/Q Imbalance block.

4. **Automatic Gain Control:** The role of the AGC circuit is to provide relatively constant output amplitude so that circuits following the AGC circuit require less dynamic range. If the signal level changes are much slower than the information rate contained in the signal, then an AGC circuit can be used to provide a signal with a well defined average level to downstream circuits. It is an adaptive system whose average output signal level is fed back to adjust the gain to an appropriate level for a range of input signal levels. For example, without AGC the sound emitted from an AM radio receiver would vary to an extreme extent from a weak to a strong signal; the AGC effectively reduces the volume if the signal is strong and raises it when it is weaker. AGC algorithms often use a PID controller where the P term is driven by the error between expected and actual output amplitude.

5. **Phase and frequency error compensation:** Rotates the signal to represent correction of phase and Doppler error on the link. This block simply corrects using the same values as the Phase / Frequency Offset block.

6. **Raised Cosine Receive Filter:** This block applies a matched filter to the modulated signal using the square root raised cosine pulse shape.

   **Functional block Parameters -**
   Filter type: Square root
   Roll off factor: 0 .2
   Down sampling factor: 8

7. **Rectangular QAM Demodulator Baseband:** The Rectangular QAM Demodulator Baseband block demodulates a signal that was modulated using quadrature amplitude modulation with a constellation on a rectangular lattice. The demodulator algorithm maps received input signal constellation values to M-ary integer I and Q symbol indices between 0 and $\sqrt{M} - 1$ then maps these demodulated symbol indices to formatted output values.
   **Functional block parameters**
   - M-ary number: 16
   - Constellation ordering : Gray
   - Minimum distance (The distance between two nearest constellation points):  2
   - Decision type : Hard decision

8. **Sink:** The sink is the BER estimate section, which compares a transmitted data stream with a receive data stream to calculate the Bit error rate of a system. It also outputs the number of error events that have occurred, and the total number of bits or symbols compared.

### III. Modelling system Impairments.

The transponder is a central element in the end-to-end communication link and is one of the elements in overall performance. There the transmitting earth –station on the up-link side will cause its share of distortion, as will the receiving earth station on the downlink side. To obtain maximum power output with the highest efficiency, the amplifier should be operated at its saturation point, which produce, AM / AM   and AM / PM conversation. The next significant impairments to digital transmission are from the filters, which constraints bandwidth and introduce delay distortion. Simple link budgeting techniques are available for evaluating links with noise.

#### *3.1 Link Budget*

Link budget is a tabular method for evaluating the received power and noise power in a radiofrequency link. Table 1 shows the typical link budget for an extended C-band downlink using a global beam on a GEO satellite and a 7.2-m station antenna. The link budget must be calculated for individual transponders, and must be repeated for each individual links.

Table 1 gives the data's of KPTCL which is used in developing the model.

| | |
|---|---|
| Earth station Transmitter antenna Gain | 52.48db |
| Satellite Transmitter antenna gain | 31db |
| Earth station Receiver antenna gain | 36.85db |
| Satellite Receiver antenna gain | 38.2db |
| Transponder bandwidth | 36MHz |
| Up-link frequency band | 6.875 - 6.9465 GHz |
| Downlink frequency band | 4.650 - 4.7215 GHz |
| Up-link loss  and   Down-link loss | 221db   and   217db |

#### **3.2 Link budget calculations**

The link between the satellite and Earth station is governed by the basic microwave radio link equation:

$$P_r = \frac{P_t \, G_t \, G_r \, C^2}{4\pi^2 \, R^2 \, f^2} \qquad (1)$$

Where $P_r$ is power received by the receiving antenna: $P_t$ is the power applied to the transmitting antenna: $G_t$ is the gain of the transmitting antenna, $G_r$ is gain of the receiving antenna, C is the speed of light (c = 3 x 10$^8$ m/s): R is the range (path length) in meters: and f is the frequency in hertz .Almost all link calculations are performed after converting from products and ratios to decibels. This uses the unit popular unit of decibels, thus converting the equation (1) into decibels. It has the form of a power balance as $P_r = P_t + G_t + G_r - \text{Path-loss}$

All link budgets require knowledge of the free space path loss between the earth station and the satellite and the noise powers in the operating bandwidth.

Free space Path loss: $L_p = 20 \log (4\pi R / \lambda)$ (2)

The **Parameters:**

- Distance = 37000 km
- Uplink Frequency = 6946 MHz
- Downlink frequency = 4721 MHz

**Gain (Tx and Rx. Dish Antenna Gain):** Gain of the antenna is given by -

$$G = \frac{4\pi \eta A}{\lambda^2} = \eta (\pi D/\lambda)^2 \qquad (3)$$

Where $\eta$ is the antenna efficiency, A is the effective area, $A = \pi r^2$ and $\lambda$ is the wavelength.

Table 2 and Table 3 helps in the calculation of Transmitter and Receiver Antenna Gains

**Table 1:** Transmitting antenna parameters

| Antenna size | 7.2m |
|---|---|
| Uplink frequency | 6946Mhz |
| Antenna efficiency | 64% |
| Pointing losses | 0.5 dB |

From the parameters given in Table 1, **Gain $G_u$ = 52.48 dB**

The maximum VSAT Antenna Gain for uplink was found by taking the maximum carrier frequency into account as per Table 1. And VSAT Antenna Gain for downlink was obtained by taking the maximum carrier frequency as shown below.

**Table 2:** Receiving antenna parameters

| Antenna size | 1.8m |
|---|---|
| Downlink frequency | 4721Mhz |
| Antenna efficiency | 63% |
| Pointing losses | 0.5 dB |

From the parameters given in Table 2, **Gain = 36.85 dB**

## IV. Result and Discussion

All signal sources in the signal processing and communication's can generate frame based data. In this work, the signal is frame based and samples are propagated through a model and multiple samples are processed in batches. Frame – based processing takes advantage of Simulink matrix processing capabilities to reduce overhead. Complex modulation scheme are best viewed using a scatter diagram. The scatter diagram allows us to visualize the real and imaginary (in-phase and quadrature) component of the complex signal. Thus Fig 7 shows the Scatter plot of VSAT Earth station system in the transmitter and Fig.8 shows the Scatter plot of VSAT _ satellite transmitter constellation showing the effect of phase offset. Fig 9 shows the VSAT _Satellite receiver constellation showing the effect of frequency offset, and fig 10 shows VSAT _satellite received constellation after RF corrected Impairments.

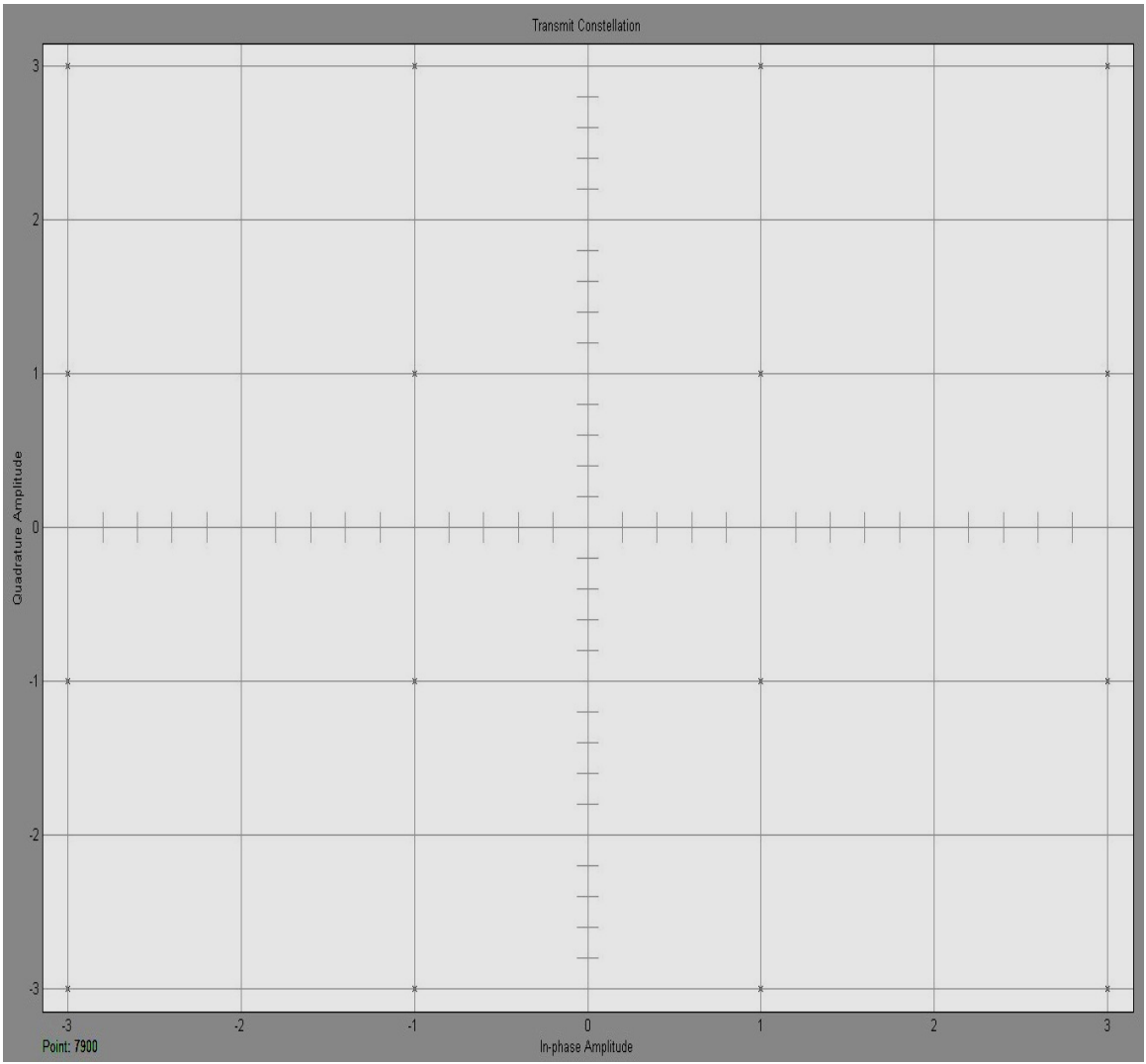

FIG 7: VSAT Earth station transmitter Constellation diagram

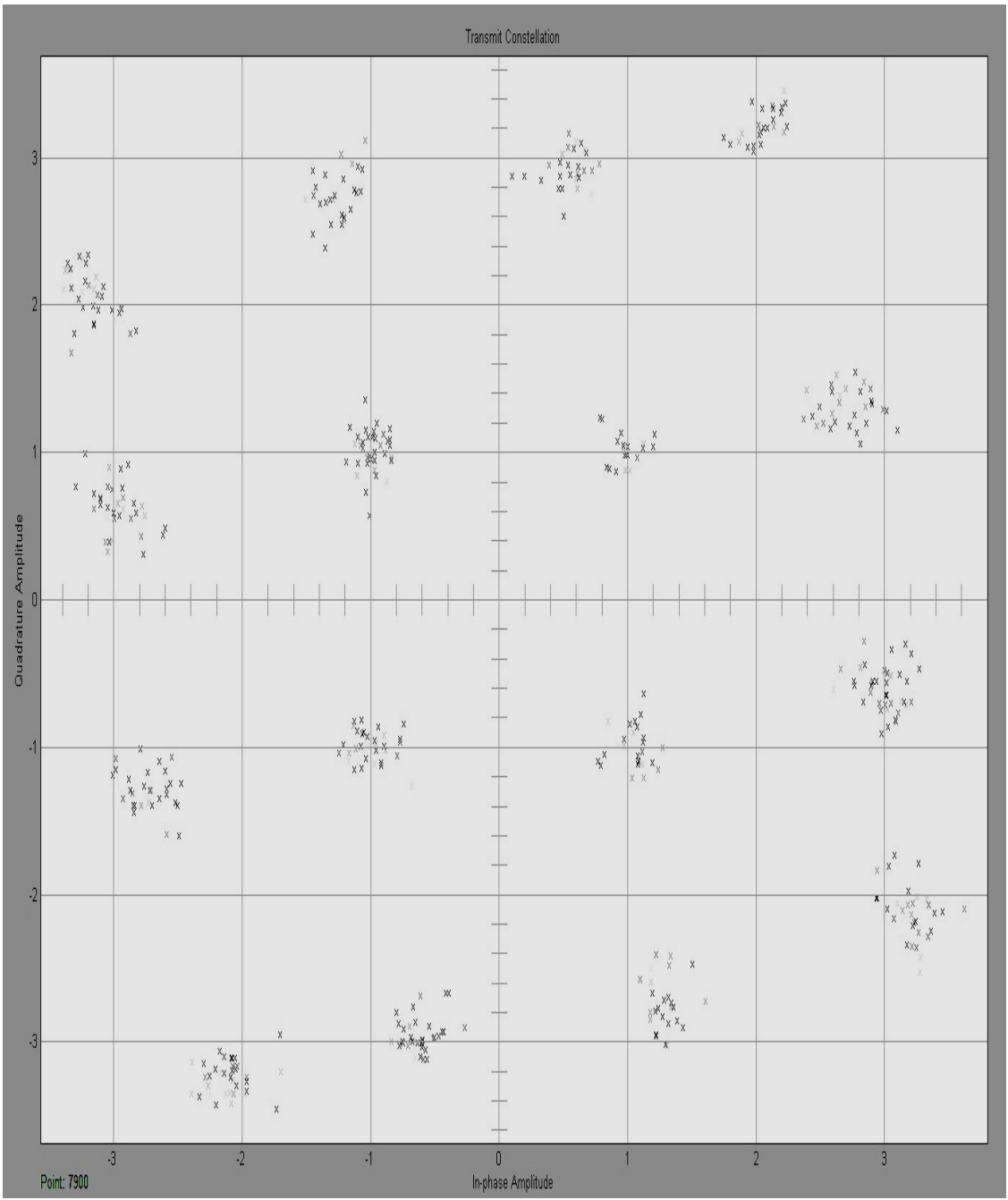

**FIG8:** VSAT-satellite transmitter constellation showing the effect of phase offset

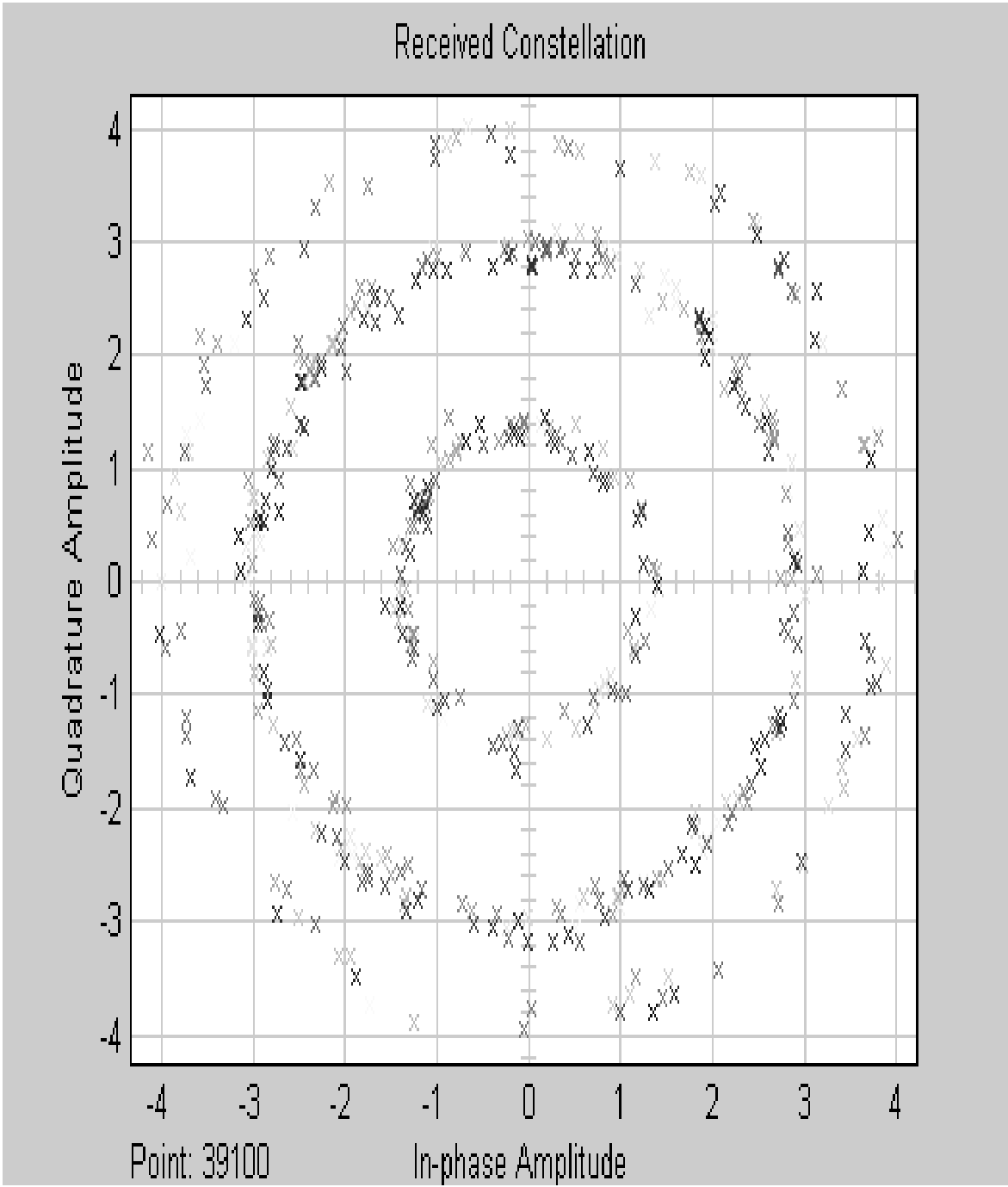

**FIG 9:** VSAT_satellite received constellation showing the effect of frequency offset

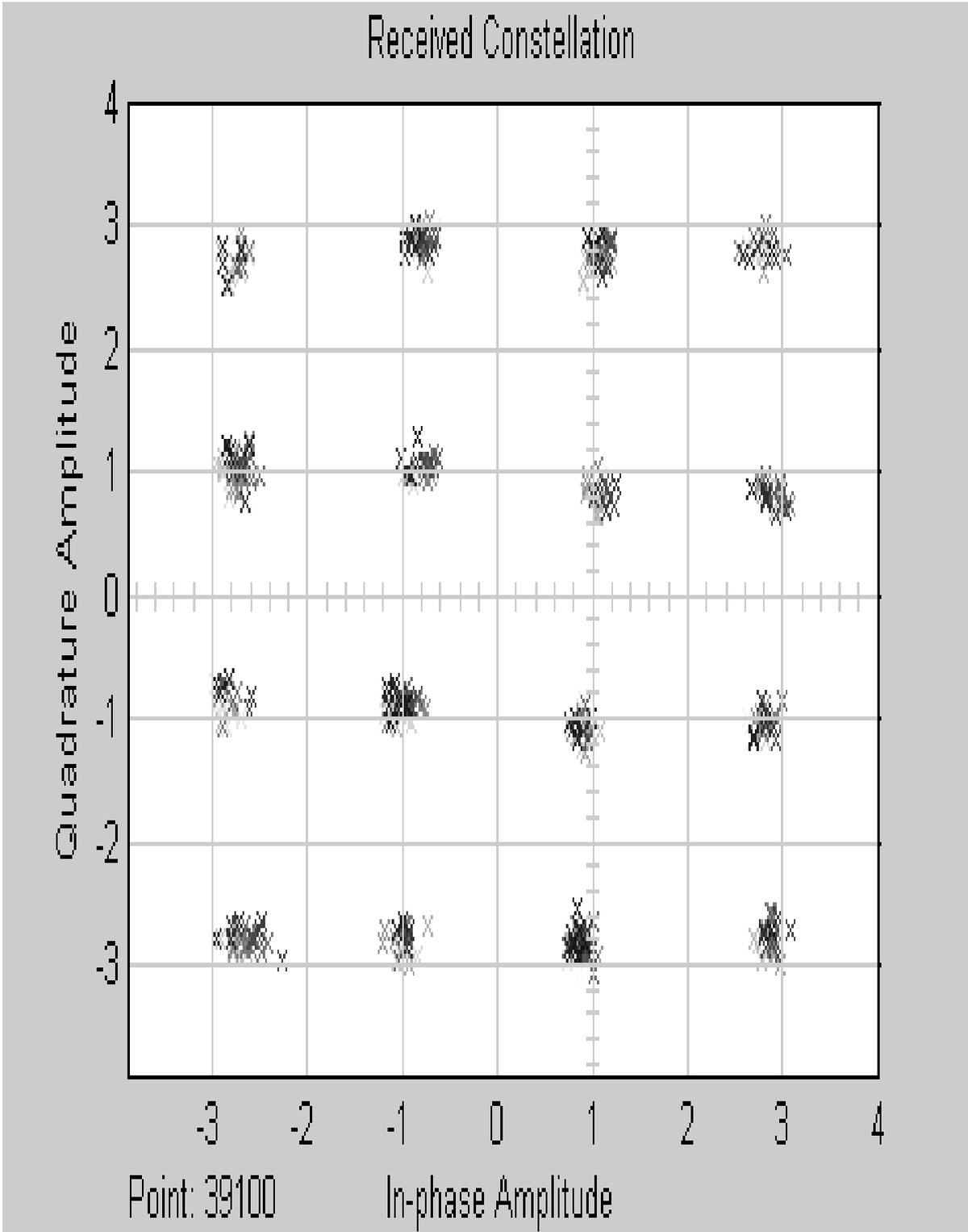

**FIG 10:** VSAT _ satellite received constellation after phase/frequency correction

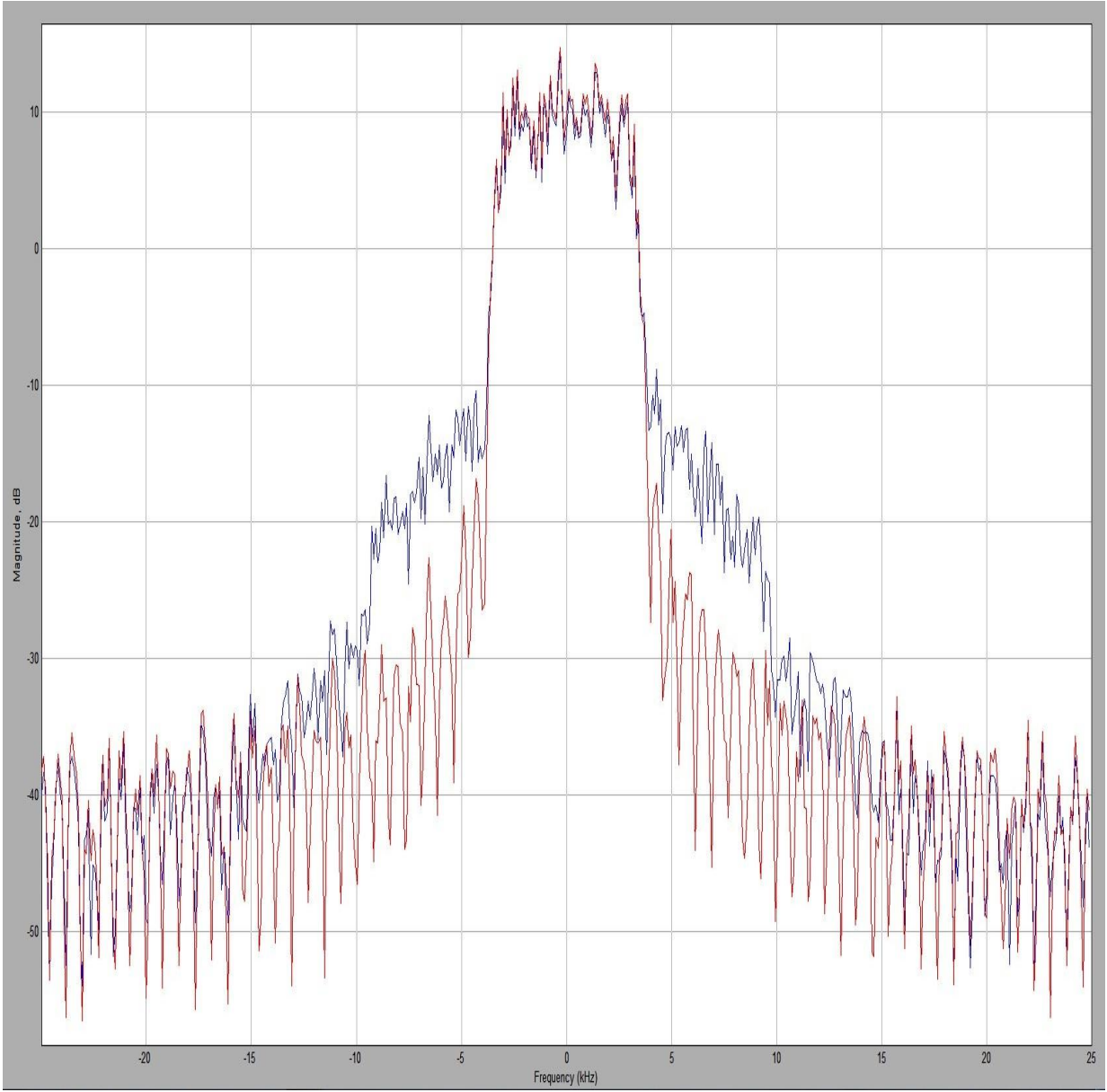

**FIG 11:** VSAT transmitter and receiver signal power spectrum.

**Spectrum diagram**

Fig11 shows the spectrum of the modulated / transmitted signal as indicated in red colour and the received signal as shown in blue colour. Both spectrums are almost similar but some effects of thermal noise caused by the Receiver thermal noise block can be seen in the receiver signal Spectrum.

**BER Estimation**

Bit Error Rate calculator block compares the Data transmitted message with the received data message and displays the error as BER as shown below.

BER = 0.1236 (increased due to phase effect)

BER = 0.5001 (increased due to frequency effect)

To overcome these two BER values, a compensation of phase /frequency offset is added in the receiver section and the constellation of such correction is visualized in Fig.10. The final BER values after performing phase/frequency compensation are obtained as shown below.

BER= 0.00052 (after performing phase/frequency offset compensation)

# Conclusion

The results can be concluded by comparing the three analysis of simulated results.
The first analysis of the case study exhibits the effect of Phase/Frequency offset whose values are high without compensation block.

BER = 0.1236 (increased due to phase effect)

BER = 0.5001(increased due to frequency effect).

Fig. 8 shows the constellation diagram at the Receiver before phase and frequency correction. Where there is a phase tilt of 15 degrees and fig. 9 shows the effect of frequency of 2 Hz. These distorted signals can be corrected by using phase/frequency compensation network.

Thus the second analysis of the case study exhibits the BER = 0.00052. whose value is desired by considering the effect of phase/frequency offset after passing through satellite Transponder. As shown in fig 10.
.
The third analysis shows receiver spectrum diagram which gives the effect of thermal noise as indicated by two colours, red colour indicates transmitted signal and blue colour caused by receiver thermal noise block.

Thus Table 1 to 3 gives the satellite parameters of load dispatching centre of Karnataka Power system at Bangalore. Where the same values have been implemented in the model to simulate the BER results whose values exceeds $10^{-3}$ in analysis 1 and the values can be further corrected by using compensation circuit.

## Acknowledgments

The authors are very grateful to the Management of Vidya Vardhaka College of Engineering, Mysore, Karnataka, India, The National Institute of Engineering, Mysore, Karnataka, India, Sapthagiri College of Engineering, Bangalore, Karnataka, India for their constant encouragement, and motivation during their work.

**Authors**

T.P. Surekha, received the B.E degree from Mangalore University, Mangalore. And M.Tech degree from Visvesvaraya Technological University, Belgaum. Presently she is working as Associate Professor in the department of Electronics and Communication Engineering, Vidyavardhaka College of Engineering, Mysore. Her research interest include Power-line communication System, Automation and Simulation of Communication System in Power System. Modelling and Simulation of Communication systems.

T. Ananthapadmanabha received the B.E. degree, M.Tech degree and Ph.D. degree from the University of Mysore, Mysore. Presently, he is working as Professor and Head in the Department of Electrical Engineering, The National Institute of Engineering, Mysore. He is also Honorary Secretary of Institute of Engineers, (India), Mysore Local Centre, Mysore. His research interest include voltage stability, distribution automation and AI applications in power system. Simulation of Power system and Communication Systems.

C Puttamadappa, received the B.E degree from Mysore University, Mysore. M. E degree from Bangalore University, Bangalore and Ph.D degree from Jadvapur University, Kolkatta. Presently he is working as professor in the department of Electronics and Communication Engineering, Sapthagiri College of Engineering, Bangalore, His research interest include Power Electronics, Simulation of communication systems, and Mobile Wireless Networks.